\documentclass{sig-alternate}

\usepackage{comment}

\usepackage{graphicx}
\usepackage{epstopdf}
\usepackage[usenames,dvipsnames]{xcolor}

\usepackage{amsmath}
\usepackage{amsfonts}
\usepackage{amssymb}
\usepackage{url} 
\usepackage{multicol}
\usepackage{multirow}
\usepackage{xspace}
\usepackage{listings}
\usepackage{array}
\usepackage{booktabs}
\usepackage{threeparttable}
\usepackage[utf8x]{inputenc}
\usepackage{balance}
\usepackage[T1]{fontenc}
\usepackage[%
	bookmarks, colorlinks, citecolor=green, urlcolor=blue,
	filecolor=blue, linkcolor=blue, pdfpagelabels=false%
]{hyperref}

\usepackage[scaled]{helvet}
\usepackage{enumitem}
\usepackage{courier}            
\usepackage[scaled]{helvet} 
\usepackage{url}                  
\usepackage{listings}          
\usepackage{enumitem}      
\usepackage{pifont}

\usepackage[ruled,lined,linesnumbered,vlined]{algorithm2e}

\usepackage{xspace}

\usepackage{tikz}
\usetikzlibrary{arrows,positioning,calc,shapes.geometric,shapes.symbols,shapes.arrows}
\usepackage{standalone}

\usepackage{contour}

\newcommand{\eg}{\hbox{\emph{e.g.}}\xspace}
\newcommand{\ie}{\hbox{\emph{i.e.}}\xspace}

\makeatletter
\def\@copyrightspace{\relax}
\makeatother

\begin{document}

\title{Toward Rapid Transformation of Ideas into Software}

\numberofauthors{1}
\author{
\alignauthor Mehrdad Afshari\qquad\qquad Zhendong Su\\[1pc]
\affaddr{Department of Computer Science, University of California, Davis}\\
\email{\{mafshari, su\}@ucdavis.edu}
}

\maketitle

\begin{abstract}

A key mission of computer science is to enable people realize their
creative ideas as naturally and painlessly as possible. Software
engineering is at the center of this mission --- software technologies
enable reification of ideas into working systems.  As computers become
ubiquitous, both in availability and the aspects of human lives they
touch, the quantity and diversity of ideas also rapidly grow.  Our
programming systems and technologies need to evolve to make this
reification process --- \emph{transforming ideas to software} --- as
quick and accessible as possible.

The goal of this paper is twofold. First, it advocates and highlights
the ``transforming ideas to software'' mission as a moonshot for
software engineering research. This is a long-term direction for the
community, and there is no silver bullet that can get us there.  To
make this mission a reality, as a community, we need to improve the
status quo across many dimensions. Thus, the second goal is to outline
a number of directions to modernize our contemporary programming
technologies for decades to come, describe work that has been
undertaken along those vectors, and pinpoint critical challenges.

\end{abstract}

\section{Introduction}
\label{sec:intro}

There has been no shortage of creative technological ideas, but few
have been realized --- it is a daunting task to transform an idea into
a working prototype. Indeed, software engineering --- the process of
expressing and refining ideas in a programming language --- has been
regarded one of the most challenging human endeavors. Programming
innovations, such as procedural abstraction and object orientation,
have helped increase programmer productivity.  However, we still build
software essentially the same way as we did decades ago. As a
community, we should rethink and redesign methodologies and techniques
for programming to make software development more natural and painless
to help people realize their creative ideas.

We believe that \emph{transforming ideas into software (TIIS)} should
be identified as a long-term, catalytic mission for the software
engineering community. Decades of research and development have led to
better languages, methodologies, tools, environments, and processes.
However, it is fair to say that most have been incremental
improvements and do not promise significant advances demanded for the
mission. Identifying and highlighting the TIIS mission can help unite
the community and clarify important research focuses to achieve
significant innovations. 

The TIIS mission requires a multi-faceted approach, which we organize
around several key principles:
\begin{itemize}
\item \emph{Quick experimentation}: to provide developers with
  immediate feedback on their code modifications and allow them to
  experiment with incomplete systems; 
\item \emph{Programming knowledge reuse}: to allow developers quick
  access to the vast amount of accumulated programming knowledge and
  wisdom; 
\item \emph{Proactive programming assistant}: to monitor the
  developers' actions and proactively feed them relevant information
  about the program; and
\item \emph{Intelligent, conversational interfaces}: to provide
  alternative interfaces that allow developers to express their
  intentions and conduct interactive exchanges with the system. 
\end{itemize} 

The two core questions in programming are ``What'' and ``How'': (1)
``What'' specifies the intention, and (2) ``How'' concerns the
solution. The first three principles center around the ``How''
question, while the last principle the ``What''. Next, we discuss the
above principles, and pinpoint specific research problems and
challenges.

\section{Directions and Challenges}
\label{sec:idea}

The vision for quick transformation of ideas into software is broad, and
advances in a number of directions are necessary and can move
the state of affairs forward.  We discuss several directions
that we have identified that can be influential toward our
goal.  We have done early work along some of these directions 
and hope the community as a whole can help accelerate the progress 
toward improving programming and in particular, the pace of concretizing ideas.

\subsection{Quick Experimentation}

Live programming has gained momentum following Bret Victor's
presentation~\cite{bretvictor:talk2012}, in which he highlighted
the importance of immediate connection between the idea
and observing its effect, not just as a catalyst, but as an
enabler, in an effective creative process.  Since then,
several live programming environments, \eg Xcode~\cite{xcode}
(via its Playground feature) and LightTable~\cite{lighttable}
that have been influenced by this principle.

Prorogued programming~\cite{prorogue:onward2012} is a programming
paradigm that explicitly deals with the issue of quick
experimentation.  It is focused on liberating the programmer from
having to deal with programming concerns that are necessary to get
a partial, incomplete, program running and meaningfully experiment
with it and observe its behavior.  It does so by providing the
ability to annotate function calls or type instantiations
with a special keyword, \texttt{prorogue}.  The prorogue keyword
acts as a hint for the compiler to let it know that the implementation
for the particular method being called is unavailable.
At runtime, after a prorogued call is executed, a lazy
\emph{future} object is returned in lieu of the return value and the
program execution continues.  Later, if the value of that object is
consulted during the program execution, the user will be asked to
provide a concrete return value for the call interactively,
while presenting him the actual arguments in that specific invocation.
The user interaction will then be recorded and persisted for the rest
of the program execution and for subsequent runs, so that the program
can be run and experimented with in spite of the unimplemented method body.

In effect, prorogued programming aids quick experimentation and
top-down design by letting the programmer freely rearrange his workflow
as he sees fit, rather than having to follow an order imposed by
the toolchain they are using.

More interestingly, through \emph{hybrid computation}, prorogued calls can act
as hooks to glue a program written in an imperative textual programming
language into more domain-specific programming systems that would capture the
human intent much better and in a more concise fashion for particular
purposes.  The other end of hybrid computation does not even have to be an
imperative program.  It can be a machine learning model that is trained to
provide the desired function that would be hard to express the host language.
Alternatively, it can be an interactive system that computes the desired output
through some user interaction.  It is possible to have a hybrid computation
engine that is mostly similar to mainstream textual programming languages,
except it is much \emph{softer} when it comes to interpreting programmer
intent, leaving room for the compiler to make educated guesses and at the same
time be more lenient to programmer mistakes, at the expense of precision.

\subsection{Programming Knowledge Reuse}

Software is rarely written from scratch.  Rather, programs are generally
composed of smaller pieces.  That makes software engineering activity largely a
system integration process.  Software engineers build more complex abstractions
out of simpler ones and that lets them build increasingly sophisticated
systems.  While seeing the effects of a program live helps, the question
remains that given that there are vast amounts of source code available on the
Internet, should we move from writing new code to casting programming as a
search problem?

The programming knowledge publicly available today 
comes in various forms, such as questions and answers on Stack
Overflow~\cite{stackoverflow}, sometimes including code snippets
as well as answers, or through publicly accessible code repositories
such as the ones hosted on GitHub~\cite{github}.

Commercial software development endeavors also collect internal
data about their development process, including the version history
of the code base, data about bugs and defects, and free-form
knowledge in form of comments written on the code review tool,
wikis, and sometimes in other forms, like tracking the time the
programmers spend on various tasks, storing the search queries
they perform~\cite{caitlin:codesearch2015}, or looking at their
behavior within the development environment.

In software engineering practice, major effort is expended
to integrate various systems and assemble a program from building
blocks.  Given the large amount of code available, it is
conceivable that what a programmer plans to write is already
written and available in some shape or form.
Effective code search can help the programmer discover the
existing functionality from existing code bases
import it in the code being written~\cite{bingcodesearch}.

With a mechanism to locate pieces of functionality 
through existing APIs or code snippets mined from the Internet,
we need to be able to run the resulting \emph{mashup} consisting of
the different pieces and quickly experiment with them.
A programming paradigm like prorogued programming is well-suited
for this task.  Proroguing programming concerns not only helps
in piecing together the building blocks of functionality discovered
in the existing code bases, but also provides a way to effectively
insert \emph{holes} in the program, which can be filled later.
Filling these holes can be done through traditional implementation,
\ie writing a body for the unimplemented method, or it can be done
through more innovative means, like acting as a signal in addition
to the search query and helping the search engine know the context
in which the code snippet being searched for is going to be live in.
In addition to providing that context, the input/output examples
persisted during runtime invocation of prorogued calls are a great
source of input for an I/O-based code search engine and act as a
final filter for validation of code found by a simple keyword
based code search engine.

Collecting data about the programmer's actions is helpful in other ways
as well.  By looking at the actions the programmer performs within their
development environment, for example, it is possible to predict what they
intend to accomplish and propose shortcuts to achieve what
they are aiming for more
efficiently~\cite{ideplusplus,Murphy-Hill:recommending:fse12}, thereby
educating the programmer and making them more effective in the future.
Obviously, this can help the IDE designer improve the development
environment and simplify its user interface as well.

Reusing programming knowledge is also beneficial in activities
beyond writing code.  For instance, we are able to leverage
debugging knowledge accumulated over the previous debugging
sessions to automatically help the programmer fix the new, similar,
issues~\cite{oscilloscope:oopsla2012}.  One way that has been
accomplished is by collecting and matching the program traces
that exhibit buggy behavior and pattern matching new traces
against the ones in the bug database, revealing information
about the nature of the bug and how it was previously fixed,
potentially helping the programmer understand and fix the new issue.

\subsection{Proactive Programming Assistant}

Many programming analysis tools have been developed.  In practice,
program analyses are primarily left to compile time and later.  We
believe that we should surface as much relevant information as
possible to the programmer as soon as possible.  Programming tools
should capture runtime data and run background static or dynamic
analysis while the code is being written, and guide the programmer
throughout the coding process.  With the popularity of
compiler-as-a-library solutions like libclang~\cite{clangtooling} or
Roslyn~\cite{roslyn}, we are already seeing this shift accelerating.
Our editors are indeed becoming more proactive in issuing compiler
warnings and providing safe refactoring tools.

That said, in particular, the potential for capturing dynamic
information and surfacing it in useful ways while coding remains largely untapped.
Among other things, the captured data can feed into the live programming
aspects of the system, providing the user with a concrete view of
the program, instead of a purely abstract one relying solely on static
analysis.  We believe what information is useful to the programmer and
how to best surface it will be an exciting and impactful avenue for
further research.

Speculative analysis~\cite{BrunHEN2010:FOSER} and its follow up work
can perhaps be viewed as a specific instance of this direction, where
the focus is on using speculative analysis in the background to help
developers make certain decisions.
 
\subsection{Human Interface Innovation}

Textual code is a precise and expressive medium for communicating
intent.  Looking back at the past half century of programming history,
it is hard to see it going away anytime soon.  However, most of the
computing devices shipped today are phones that do not have a
physical keyboard and mouse.  While it is conceivable that most of
the professional programming activity would not be done on such
devices, at least without some external accessories, it is almost
certain that end-users would want to use them to accomplish custom
computational goals or control systems by defining actions that
would happen in response to specific events.

Accomplishing this requires innovation both on the human interface
front and on the backend engine.  It is likely that many of the 
functionalities will be exposed via the artificial intelligence-based
assistant, and will be expressed as interactive voice conversations.
On the backend, we need to build more interactive programming systems
that can make educated guesses and synthesize programs with incomplete
specification, and interactively adapt it as the specification is
perfected by gradually asking for and capturing additional user input.

Moreover, even on more traditional computers, \eg desktops and laptops,
we need fundamental interface innovations to support
alternative programmers~\cite{TobyS:onward2012:recursive_drawing},
\ie people who are not professional programmers and write programs
that does computation and produces a result, which is the object
of interest to them, as opposed to the program itself.  An important
class of people who would benefit from such interface innovations
are people doing analysis on various data sets.  Already, 
tools like IPython~\cite{ipython} that have more interactive
characteristics and suit domain-specific use-cases well have gained
widespread adoption in that community.  We believe that there
is enormous potential to carry out research that would substantially
impact the life of alternative programmers in a positive way in
this area.

\section{Conclusion}
\label{sec:conc}

In this paper, we have advocated the TIIS mission for quick
realization of ideas as working systems and the modernization of our
techniques and tools to better support programming in the coming
decades.  Given the ubiquity of connected computer systems --- mostly
in the form of smartphones --- we are just at the beginning of an
explosion of ideas and applications that wait to be realized by
professional developers or end users.  Consequently, it is even more
important that we do our best as a community to improve our
programming practice to adapt it for the future challenges we will
likely face.  Achieving the TIIS mission will require significant
efforts spanning many directions.  We have identified, as a first
step, several directions centered around four principles. We hope that
the community unite to move the state-of-the-art forward toward the
TIIS vision along these and other important pertinent directions.

\bibliographystyle{abbrv}
\balance
\bibliography{refs}

\end{document}